\documentclass[a4paper, conference, 10pt, final]{IEEEtran}

\usepackage[a4paper,
            bindingoffset=0.2in,
            left=0.7in,
            right=0.7in,
            top=1in,
            bottom=2.7cm,
            footskip=.25in]{geometry}
\usepackage{enumitem}
\usepackage[nolist]{acronym} 
\usepackage{tikz}
\usepackage{bm}
\usepackage{pgfplots}
\usepgfplotslibrary{groupplots}
\usepackage{graphicx}
\usepackage{subcaption}
\usepackage{float}
\pgfplotsset{compat=newest}
\usepackage{units}
\usepackage{amsmath, amsbsy, amssymb, amsthm}
\usepackage{tikzscale}
\usetikzlibrary{arrows}
\usepackage{color}
\usepackage{epigraph}
\usepackage{circuitikz}
\usepackage{multirow}
\usepackage{booktabs}
\usepackage[plainruled]{algorithm2e}
\definecolor{darkgreen}{rgb}{0.125,0.5,0.169}
\usetikzlibrary{shapes,arrows}
\usetikzlibrary{positioning}
\usetikzlibrary{patterns}
\usetikzlibrary{decorations.pathreplacing}
\usepackage{marvosym}
\usepackage{bbm}
\usepackage{mathtools}
\usepackage{xfrac}
\usepackage{numprint}

\tikzset{>=latex}

\begin{acronym}
 \acro{CSI}{channel state information}
 \acro{UE}{user equipment}
 \acro{UL}{uplink}
 \acro{BS}{basestation}
 \acro{TDD}{time division duplex}
 \acro{FDD}{frequency division duplex}
 \acro{ECC}{error-correcting code}
 \acro{MLD}{maximum likelihood decoding}
 \acro{HDD}{hard decision decoding}
 \acro{IF}{intermediate frequency}
 \acro{RF}{radio frequency}
 \acro{SDD}{soft decision decoding}
 \acro{NND}{neural network decoding}
 \acro{CNN}{convolutional neural network}
 \acro{ML}{maximum likelihood}
 \acro{GPU}{graphical processing unit}
 \acro{BP}{belief propagation}
 \acro{LTE}{Long Term Evolution}
 \acro{BER}{bit error rate}
 \acro{SNR}{signal-to-noise-ratio}
 \acro{ReLU}{rectified linear unit}
 \acro{BPSK}{binary phase shift keying}
 \acro{QPSK}{quadrature phase shift keying}
 \acro{AWGN}{additive white Gaussian noise}
 \acro{MSE}{mean squared error}
 \acro{LLR}{log-likelihood ratio}
 \acro{MAP}{maximum a posteriori}
 \acro{NVE}{normalized validation error}
 \acro{BCE}{binary cross-entropy}
 \acro{CE}{cross-entropy}
 \acro{BLER}{block error rate}
 \acro{SQR}{signal-to-quantisation-noise-ratio}
 \acro{MIMO}{multiple-input multiple-output}
 \acro{OFDM}{orthogonal frequency division multiplex}
 \acro{RF}{radio frequency}
 \acro{LOS}{line of sight}
 \acro{NLoS}{non-line of sight}
 \acro{NMSE}{normalized mean squared error}
 \acro{CFO}{carrier frequency offset}
 \acro{SFO}{sampling frequency offset}
 \acro{IPS}{indoor positioning system}
 \acro{TRIPS}{time-reversal IPS}
 \acro{RSSI}{received signal strength indicator}
 \acro{MIMO}{multiple-input multiple-output}
 \acro{ENoB}{effective number of bits}
 \acro{AGC}{automatic gain control}
 \acro{ADC}{analog to digital converter}
 \acro{ADCs}{analog to digital converters}
 \acro{FB}{front bandpass}
 \acro{FPGA}{field programmable gate array}
 \acro{JSDM}{Joint Spatial Division and Multiplexing}
 \acro{NN}{neural network}
 \acro{IF}{intermediate frequency}
 \acro{LoS}{line-of-sight}
 \acro{NLoS}{non-line-of-sight}
 \acro{DSP}{digital signal processing}
 \acro{AFE}{analog front end}
 \acro{SQNR}{signal-to-quantisation-noise-ratio}
 \acro{SINR}{signal-to-interference-noise-ratio}
 \acro{ENoB}{effective number of bits}
 \acro{PCB}{printed circuit board}
 \acro{EVM}{error vector mangnitude}
 \acro{CDF}{cumulative distribution function}
 \acro{MRC}{maximum ratio combining}
 \acro{MRP}{maximum ratio precoding}
 \acro{MRT}{maximum ratio transmission}
 \acro{DeepL}{deep-learning}
 \acro{DL}{downlink}
 \acro{SISO}{single-input single-output}
 \acro{SGD}{stochastic gradient descent}
 \acro{CP}{cyclic prefix}
 \acro{MISO}{Multiple Input Single Output}
 \acro{LMMSE}{linear minimum mean square error}
 \acro{ZF}{zero forcing}
 \acro{USRP}{universal software radio peripheral}
 \acro{RNN}{recurrent neural network}
 \acro{GRU}{gated recurrent unit}
 \acro{LSTM}{long short-term memory}
 \acro{NTM}{neural turing machine}
 \acro{DNC}{differentiable neural computer}
 \acro{TCN}{temporal convolutional network}
 \acro{FCL}{fully connected layer}
 \acro{MANN}{memory augmented neural network}
 \acro{RNN}{recurrent neural network}
 \acro{DNN}{dense neural network}
 \acro{FIR}{finite impulse response}
 \acro{BPTT}{back-propagation through time}
 \acro{GAN}{generative adversarial network}
 \acro{ELU}{exponential linear unit}
 \acro{tanh}{hyperbolic tangent}
 \acro{BICM}{bit-interleaved coded modulation}
 \acro{OTA}{over-the-air}
 \acro{IM}{intensity modulation}
 \acro{DD}{direct detection}
 \acro{RL}{reinforcement learning}
 \acro{SDR}{software-defined radio}
 \acro{WGAN}{Wasserstein generative adversarial network}
 \acro{BMD}{bit-metric decoding}
 \acro{BMI}{bit-wise mutual information}
 \acro{LDPC}{low-density parity-check}
 \acro{IDD}{iterative demapping and decoding}
 \acro{JSD}{Jensen-Shannon divergence}
 \acro{MMSE}{minimum mean square error}
 \acro{FFT}{fast Fourier transform}
 \acro{IFFT}{inverse fast Fourier transform}
 \acro{QAM}{quadrature amplitude modulation}
 \acro{EMD}{earth mover's distance}
 \acro{TDL}{tapped delay line}
 \acro{KL}{Kullback-Leibler}
 \acro{PRACH}{physical random access channel}
 \acro{URLLC}{ultra-reliable low-latency communication}
 \acro{ANOMA}{asynchronous non-orthogonal multiple access}
 \acro{FEC}{forward error correction}
 \acro{PAPR}{peak-to-average power ratio}
 \acro{APP}{a posteriori probability}
 \acro{COTS}{commercial off-the-shelf}
 \acro{PLL}{phase locked loop}
 \acro{STO}{sampling time offset}
 \acro{SFO}{sampling frequency offset}
 \acro{CFO}{carrier frequency offset}
 \acro{CPO}{carrier phase offset}
 \acro{CSI}{channel state information}
 \acro{GNSS}{global navigation satellite system}
 \acro{ELAA}{extremely large aperture array}
 \acro{UE}{user equipment}
 \acro{DICHASUS}{\underline{Di}stributed \underline{Cha}nnel \underline{S}ounder by \underline{U}niversity of \underline{S}tuttgart}
 \acro{JCaS}{Joint Communication and Sensing}
 \acro{AoA}{angle of arrival}
 \acro{REFTX}{reference transmitter}
 \acro{MOBTX}{mobile transmitter}
 \acro{AARX}{antenna array receiver}
 \acro{SPI}{Serial Peripheral Interface}
 \acro{LO}{local oscillator}
 \acro{LLTF}{Legacy Long Training Field}
 \acro{HTLTF}{High Throughput Long Training Field}
 \acro{USB}{Universal Serial Bus}
 \acro{VCO}{voltage controlled oscillator}
 \acro{MUSIC}{MUltiple SIgnal Classification}
\end{acronym}

\definecolor{mittelblau}{RGB}{0, 126, 198}
\definecolor{violettblau}{cmyk}{0.9, 0.6, 0, 0}
\definecolor{rot}{RGB}{238, 28 35}
\definecolor{apfelgruen}{RGB}{140, 198, 62}
\definecolor{gelb}{RGB}{1, 221, 0}
\definecolor{orange}{RGB}{244, 111, 33}
\definecolor{pink}{RGB}{237, 0, 140}
\definecolor{lila}{RGB}{128, 10, 145}
\definecolor{hellgrau}{RGB}{224, 224, 224}
\definecolor{mittelgrau}{RGB}{128, 128, 128}
\definecolor{dunkelgrau}{RGB}{80,80,80}
\definecolor{anthrazit}{RGB}{19, 31, 31}

\definecolor{bgorange}{HTML}{fcc0a7}
\definecolor{bggreen}{HTML}{ccebb9}

\DeclareMathOperator*{\argmax}{arg\,max}
\DeclareMathOperator*{\argmin}{arg\,min}

\IEEEoverridecommandlockouts

\begin{document}

\title{ESPARGOS: An Ultra Low-Cost, Realtime-Capable Multi-Antenna WiFi Channel Sounder}

\author{\IEEEauthorblockN{Florian Euchner, Tim Schneider, Marc Gauger, Stephan ten Brink \\}

\IEEEauthorblockA{
Institute of Telecommunications, Pfaffenwaldring 47, University of  Stuttgart, 70569 Stuttgart, Germany \\ \{euchner,gauger,tenbrink\}@inue.uni-stuttgart.de
}

}

\maketitle

\begin{abstract}
    Multi-antenna channel sounding is a technique for measuring the propagation characteristics of electromagnetic waves that is commonly employed for parameterizing channel models.
    Channel sounders are usually custom-built from many Software Defined Radio receivers, making them expensive to procure and difficult to operate, which constrains the set of users to a few specialized scientific institutions and industrial research laboratories.
    Recent developments in Joint Communications and Sensing (JCaS) extend the possible uses of channel data to applications like human activity recognition, human presence detection, user localization and wireless Channel Charting, all of which are of great interest to security researchers, experts in industrial automation and others.
    However, due to a lack of affordable, easy-to-use and commercially available multi-antenna channel sounders, those scientific communities can be hindered by their lack of access to wireless channel measurements.
    To lower the barrier to entry for channel sounding, we develop an ultra low-cost measurement hardware platform based on mass-produced WiFi chips, which is easily affordable to research groups and even hobbyists.
\end{abstract}

\acresetall

\section{Introduction}
Spatial diversity is a crucial concept in modern wireless communication systems enabling improvements in spectral efficiency.
Massive \ac{MIMO} base stations, which exploit this concept, need to continuously estimate the wireless channel to be able to communicate with \ac{UE} devices, generating large amounts of so-called \ac{CSI} as a byproduct.
Instead of discarding \ac{CSI} after a successful communication sequence, it can also be exploited by various sensing methods summarized under the umbrella term \ac{JCaS}.
While the idea of \ac{JCaS} is not new \cite{sturm2011waveform}, better spatial diversity as well as increasing spectral bandwidths are enabling additional sensing applications which rely on high-dimensional \ac{CSI} such as Wireless Channel Charting \cite{studer}.

To better understand performance characteristics of massive \ac{MIMO} systems and to develop \ac{JCaS} algorithms, several universities and research laboratories are operating so-called channel sounders, i.e., devices specifically built for \ac{CSI} measurements.
Typically, those channel sounders use protocols and waveforms specifically designed to accomodate for channel estimation.
While the acquired channel data is invaluable to progress in the field of wireless communication systems, channel sounders are usually not suitable for use outside of this field of research:
They require qualified operators, are expensive and difficult to set up and may even involve obtaining frequency allocations or an amateur radio license.

Due to these downsides, researchers in other domains who work with wireless channel data have often turned to WiFi:
Often-quoted WiFi sensing applications include human presence and activity detection, traffic monitoring, gesture and sign language recognition, localization, imaging and others \cite{ma2019wifi}, with many privacy-sensitive applications such as gesture / pose detection \cite{adib2013see} \cite{geng2022densepose} also sparking interest and concerns among the general public.
Common methods for WiFi \ac{CSI} acquisition are the use of \acp{SDR}, software libraries for Intel and Atheros WiFi cards \cite{Halperin_csitool} \cite{atheroscsi} as well as software libraries for extracting CSI from WiFi-capable Espressif ESP32 microcontrollers \cite{espcsi} \cite{wiesp}.
The major disadvantage of these inexpensive channel sounding techniques is the low number of \ac{MIMO} channels, usually only three antennas or even fewer, which is unfavorable for sensing applications that greatly benefit from spatial diversity.
Simply using multiple WiFi cards or microcontrollers without taking additional precautions does not produce phase-coherent \ac{CSI} due to lack of synchronization -- data acquired this way would be unusable for many \ac{JCaS} algorithms, including e.g. \ac{AoA} estimation.
Proposed workarounds like antenna switching \cite{XieSWAN} come with their own downsides, such as requiring multiple frame transmissions for one sensing operation.

\begin{figure}
    \centering
    \vspace{-0.3cm}
    \includegraphics[width=0.9\columnwidth]{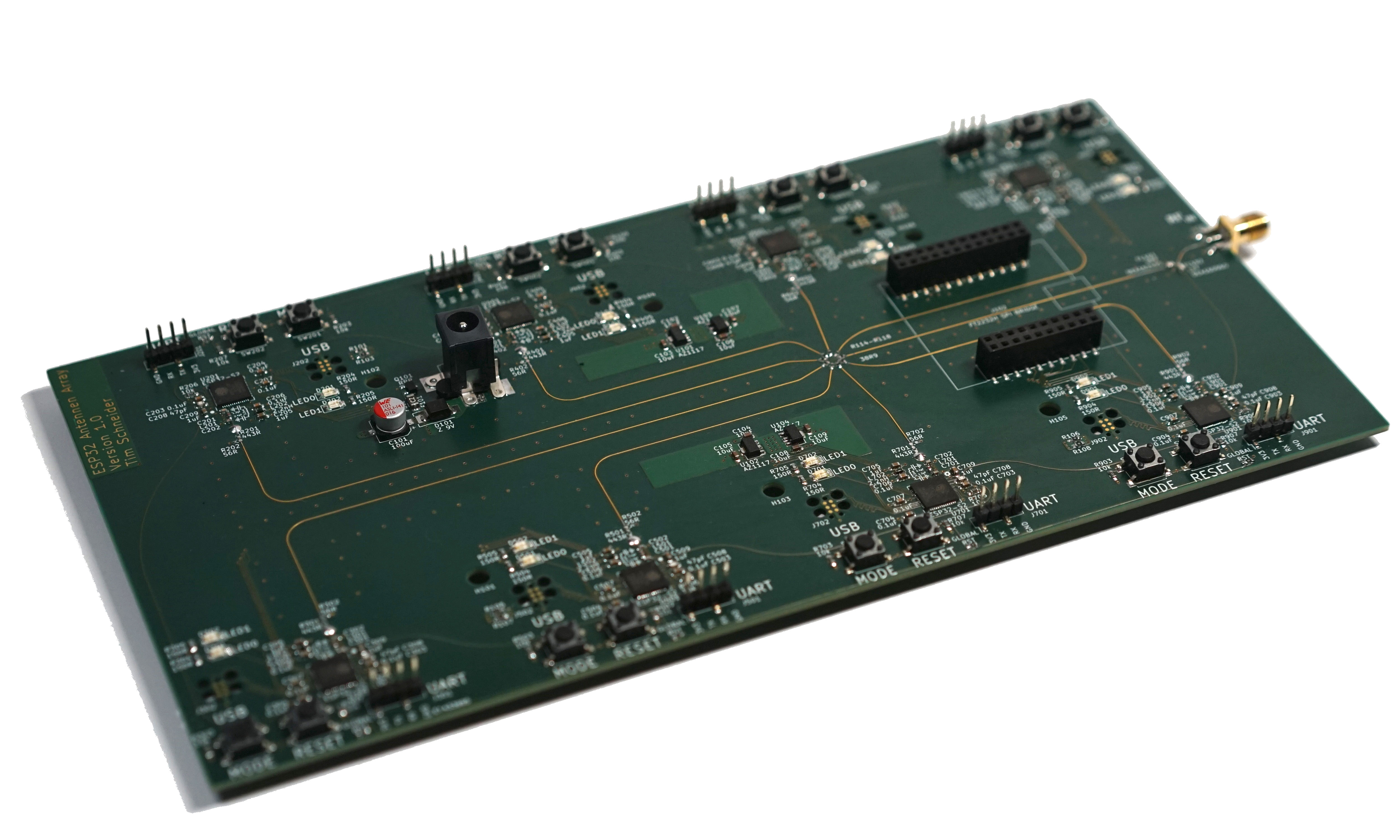}
    \vspace{-0.1cm}
    \caption{Picture of prototype circuit board for ESPARGOS}
    \label{fig:prototypepcb}
    \vspace{-0.5cm}
\end{figure}

To overcome these challenges, we propose ESPARGOS (a portmanteau of ESP32 and Argos Panoptes, the greek mythological many-eyed giant), a low-cost, single-circuit board (see Fig. \ref{fig:prototypepcb}) approach to a coherent and easy-to-operate many-antenna WiFi channel sounder, geared towards prototyping WiFi sensing devices.

The remainder of this paper is structured as follows: The hardware and software architecture of ESPARGOS are explained in Section \ref{sec:hardware} and \ref{sec:software}, respectively. Characterizations and measurement results are presented in Section \ref{sec:measurements} and Section \ref{sec:conclusion} summarizes our work and provides an outlook.

\section{Hardware Architecture}
\label{sec:hardware}

\begin{figure}
    \centering
    \includegraphics[width=\columnwidth]{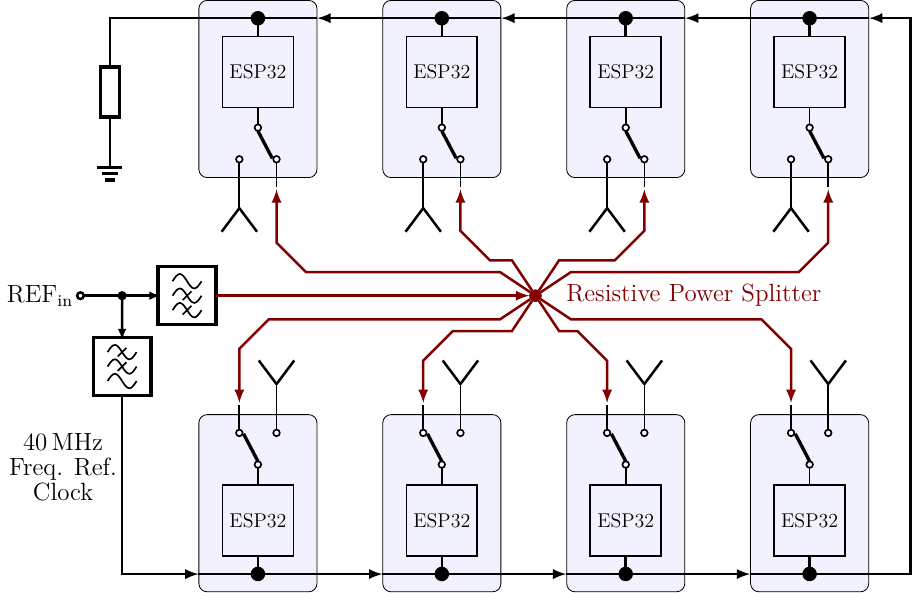}
    \caption{Block diagram showing all components related to synchronization on an ESPARGOS circuit board, with the distribution network for the phase reference signal in red.}
    \label{fig:pcbblockdiagram}
\end{figure}

Inspired by KrakenSDR \cite{krakenrf}, ESPARGOS is made up of several ESP32 microcontrollers, operated in receive-only mode.
All hardware components, including eight ceramic antennas, are mounted on a single four-layer FR4 circuit board, making ESPARGOS inexpensive and easy to produce and deploy.
A pin header on the circuit board exposes a \ac{SPI} bus, which can be used to interface the board with a computer.
As illustrated in Fig. \ref{fig:pcbblockdiagram}, all microcontrollers are clocked by a daisy-chained $40\,\mathrm{MHz}$ reference oscillator, ensuring frequency synchronization.

\subsection{Source of Phase Uncertainty}
\begin{figure}
    \centering
    \includegraphics[width=0.8\columnwidth]{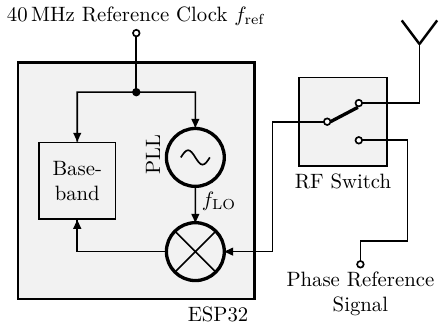}
    \caption{Block diagram of a single receiver: The ESP32 is frequency-synchronized using both a $40\,\mathrm{MHz}$ reference clock and, to ensure coherence, a WiFi-based phase reference signal}
    \label{fig:esp32rfswitch}
\end{figure}

\begin{figure}
    \centering
    \begin{circuitikz}
    \ctikzset{bipoles/thickness=2}
	\tikzset{sigblock/.style={minimum width = 1cm, minimum height = 1cm, thick, align = left, font = \small}}

	\node (xorect) at (0, 0) {$f_\mathrm{ref}$};

	\node (divider) [sigblock, draw, right = 0.5cm of xorect, font = \Large] {$\sfrac{1}{r}$};

	\node (pfd) [mixer, draw, right = 1cm of divider] {};
	\node [above = 0.1cm of pfd, align = center] {Phase\\Detector};
	\node (vco) [oscillator, draw, right = 2.5cm of pfd] {};
	\node [above = 0.1cm of vco] {VCO};
	\node (plldiv) [sigblock, draw, below = 0.7cm of vco, xshift = -2cm, font = \Large] {$\sfrac{1}{n}$};
	\node (out) [right = 0.5cm of vco] {$f_\mathrm{c}$};

	\draw [-latex] (xorect) -- (divider);
	\draw [-latex] (divider.east) to[short, l=$f_\mathrm{d}$] (pfd.west);
	\draw [-latex] (pfd.east) to[lowpass, name=lpf] (vco.west);
	\draw [-latex] (vco.south) |- (plldiv.east);
	\draw [-latex] (plldiv.west) -| (pfd.south);
	\draw [-latex] (vco.east) -- (out.west);
\end{circuitikz}
    \caption{Block diagram of a typical integer-$n$ phase-locked loop. The \ac{PLL} in the ESP32 may be implemented in a similar fashion, synthesizing carrier frequency $f_\mathrm{c}$ from reference clock $f_\mathrm{ref}$.}
    \label{fig:pll}
\end{figure}
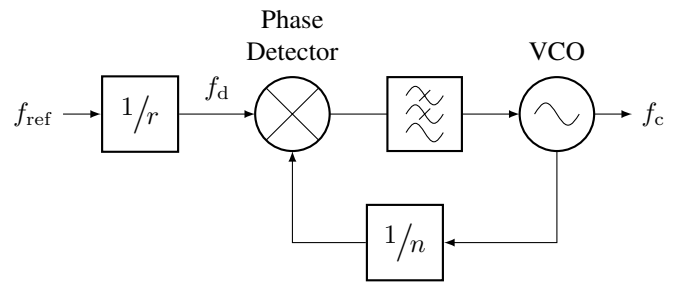

Reference clock distribution is insufficient for achieving phase coherence:
While the internal structure of the ESP32 chip is not publicly known, it is reasonable to assume that the \ac{LO} signal is generated from the reference clock using a \ac{PLL} (see Fig. \ref{fig:esp32rfswitch}), which may exhibit phase uncertainty between reference clock and output clock.
This phase uncertainty is a common property of \acp{PLL} and can have multiple sources, including ambiguities in the phase detector, unknown initial digital counter states in the forward ``$\sfrac{1}{r}$'' frequency divider (see Fig. \ref{fig:pll}) and \ac{VCO} center frequency inaccuracies \cite{brown2005method} \cite{phaseambiguity}.
By measurement it was determined that, indeed, the initial phase changes after a chip reset and after modifying the \ac{LO} frequency (WiFi channel), likely every time the \ac{PLL} has to acquire a new frequency lock.
Furthermore, even if this source of ambiguity did not exist, analog components may introduce receiver-specific phase shifts due to e.g. process variation or temperature dependencies.
Hence, an additional, periodically performed phase calibration step is vital to ensure phase coherence.

\subsection{Phase Reference Signal Distribution}
In the case of ESPARGOS, as shown in Fig. \ref{fig:pcbblockdiagram} and Fig. \ref{fig:esp32rfswitch}, each ESP32 is connected to an RF switch that can alternate between the ceramic antenna and a WiFi-based phase reference signal provided to every receiver using a distribution network made up of a resistive divider and microstrip transmission lines.
The phase reference signal generator must simply produce valid WiFi frames, which can be received by the ESP32 chip.
Since the expected phase differences between any two receivers are known for the reference signal (they are determined by the geometry of the distribution network), receiver-specific phase offsets can be calibrated for.

By feeding reference clock and phase calibration signal from external sources, multiple circuit boards can be combined into one large phase-coherent system.
To reduce the amount of cabling required, the $40\,\mathrm{MHz}$ reference clock and the WiFi-based phase reference signal are fed via the same coaxial cable and separated on the circuit board using appropriate high-pass and low-pass filters (see Fig. \ref{fig:pcbblockdiagram}).

\section{Software}
\label{sec:software}
The ESP32's WiFi driver provides \ac{CSI} estimated from the \ac{LLTF} \cite[Section 20.3.9.3.4]{80211n} and \ac{HTLTF} \cite[Section 20.3.9.4.6]{80211n}, which are a part of received WiFi frames, through a software interface.
In a so-called promiscuous mode, this \ac{CSI} acquisition can be carried out passively, i.e., the receivers can sniff WiFi frames without transmitting anything themselves.
The raw \ac{CSI} data is then transferred to a central computer through the \ac{SPI} bus, where \ac{CSI} from all receivers is combined and analyzed.

\subsection{CSI Acquisition}
\label{sec:csiacquisition}
Since the WiFi frames are received by the ESP32 microcontrollers independently from one another, they must first be clustered.
Received frames that correspond to the same transmitted frame can be identified based on MAC address, receiver timestamp (thanks to synchronized sampling clocks) and packet headers.
For each time index $t$, the matrix of true (physical) subcarrier channel coefficients is denoted by
\[
    \mathbf H[t] = \left( \mathbf h^{(1)}[t], \ldots, \mathbf h^{(N)}[t] \right) \in \mathbb C^{M \times N},
\]
where $M$ is the number of receive antennas and $N$ the number of usable \ac{OFDM} subcarriers.

In practice, $\mathbf H[t]$ cannot be measured directly, but the reported channel coefficients $r_{m}^{(n)}[t]$ are noisy due to estimation and quantization errors.
Despite these shortcomings, by assuming \ac{CSI} stationarity over short time intervals, the goal of obtaining good estimates $\hat { \mathbf H } = (\hat {\mathbf h}^{(1)}, \ldots, \hat {\mathbf h}^{(N)})$ of the true \ac{CSI} matrix is achievable by taking into account several consecutive measurements.
For the following derivations, we consider a single subcarrier $n$ and drop the subcarrier index in our notation, i.e. $\mathbf h[t] = \mathbf h^{(n)}[t]$ and $\mathbf r[t] = \mathbf r^{(n)}[t]$ for simplicity.
However, the approach can be extended to multiple subcarriers by simply applying the same technique to every subcarrier independently.

Let $\mathcal R = \left\{(t_1, m_1), (t_2, m_2), \ldots\right\}$ denote a set that contains tuples of time and antenna indices of all the channel coefficients measured within the considered interval, where $r_{m}[t]$ with $(t, m) \in \mathcal R$ is the channel coefficient for antenna $m$ measured at time instance $t$.
$r_m[t]$ is affected by the initial phase $\varphi$ of the transmitter, which is different for every time index $t$ and unknown due to unknown oscillator frequency offset and oscillator drift.
As a consequence, all channel coefficients for one particular time instance may be phase-rotated by a common factor $ \mathrm e^{j \varphi}$.
Furthermore, we assume that $\mathbf r \in \mathbb C^M$ is affected by additive i.i.d. noise $\mathbf n \in \mathbb C^M$:
\[
    \mathbf r = \mathbf h ~ \mathrm e^{j \varphi} + \mathbf n
\]
$\varphi \in [-\pi, \pi)$ and $\mathbf n$ are modelled as independent random variables.
While the distribution for $\varphi$ is unknown, we assume that $\mathbf n$ is zero-mean with $\mathrm E[ \mathbf n \mathbf n^\mathrm{H} ] = \sigma^2 \mathbf I$.

Instead of estimating $\mathbf h$ directly from multiple measurements of $\mathbf r$, we first compute the covariance matrix $\mathbf C$ of $\mathbf r$ as
\[
    \mathbf C = \mathrm E \left[ \mathbf r \mathbf r^\mathrm{H} \right] = \mathbf h \mathbf h^\mathrm{H} + \sigma^2 \mathbf I.
\]

In practice, we can only estimate the sample covariance matrix $\hat {\mathbf C}$ from the finite number of samples $r[t] = (r_1[t], \ldots, r_M[t])^\mathrm{T}, ~ t = 1 \ldots T$ measured within the considered interval:
\begin{equation}
    \hat { \mathbf C } = \frac{1}{T} \sum_{t = 1}^{T} \mathbf r[t] \mathbf r[t]^\mathrm{H}
    \label{eq:samplecovariance}
\end{equation}

In many cases, one or more of the receivers may miss a WiFi frame and therefore not provide \ac{CSI} for any particular WiFi frame, which makes (\ref{eq:samplecovariance}) unsuitable.
The underlying reasons for this loss of frames are unknown and hard to analyze without knowledge about the inner workings of the ESP32 chip.
Instead of requiring data from all $M$ antennas as in (\ref{eq:samplecovariance}), we can also estimate $\mathbf C$ from incomplete \ac{CSI} vectors:
Let $\mathcal T_{ij} = \{t ~ | ~ (t, i) \in \mathcal R, (t, j) \in \mathcal R\}$ denote the set of time indices for which channel coefficients are available from both antenna $i$ and $j$.
We obtain the sample covariance matrix $\hat {\mathbf C}$ by estimating the individual entries of the matrix as follows:
\[
    \hat C_{ij} = \frac{1}{|\mathcal T_{ij}|} \sum_{t \in \mathcal T_{ij}} r_i[t] r_j[t]^*
\]

Due to the limited number of samples, $\hat {\mathbf C}$ is only a noisy estimate of $\mathbf  C$, with noise matrix $\mathbf Z$ modelled as a zero-mean random variable:
\begin{equation}
    \hat { \mathbf C } = \mathbf C + \mathbf Z = \mathbf h \mathbf h^\mathrm{H} + \sigma^2 \mathbf I + \mathbf Z
    \label{eq:noisematrix}
\end{equation}

Since the distribution of $\mathbf Z$ is unknown, we propose to solve (\ref{eq:noisematrix}) for $\mathbf h \mathbf h^\mathrm{H}$ using a least-squares approach, where we minimize $\mathbf Z$ w.r.t. the Frobenius norm:
\begin{align}
    \hat { \mathbf h } &= \argmin_{\bm \vartheta \in \mathbb C^M} \left\lVert (\hat { \mathbf C } - \sigma^2 \mathrm I) - \bm \vartheta \bm \vartheta^\mathrm{H} \right\rVert^2_\mathrm{F} \nonumber \\
    &= \argmax_{\bm \vartheta \in \mathbb C^M} 2 \bm \vartheta^\mathrm{H} (\hat { \mathbf C } - \sigma^2 \mathbf I) \bm \vartheta - \lVert \bm \vartheta \rVert^4
    \label{eq:objectivefunction}
\end{align}

By setting the derivative w.r.t. $\bm \vartheta$ to $\mathbf 0$, we obtain
\begin{align}
    (\hat { \mathbf C } - \sigma^2 \mathbf I) ~ \bm \vartheta = \lVert \bm \vartheta \rVert^2 \bm \vartheta \nonumber \\
    \hat { \mathbf C } \bm \vartheta = \left(\lVert \bm \vartheta \rVert^2 + \sigma^2 \right) \bm \vartheta
    \label{eq:derivative}
\end{align}

which is a necessary condition for an optimum.
Clearly, (\ref{eq:derivative}) is fulfilled by all vectors $\bm \vartheta$ that are appropriately scaled eigenvectors of $\hat {\mathbf C}$.
Namely, they need to be scaled according to their corresponding eigenvalue $\lambda \geq 0$ (since $\hat {\mathbf C}$ is positive semidefinite), such that $\lVert \bm \vartheta \rVert = \sqrt{\lambda - \sigma^2}$.
From (\ref{eq:objectivefunction}) we see that the objective function is maximized if the eigenvector $\bm \vartheta$ with the largest corresponding eigenvalue $\lambda$ is chosen, i.e., the principal eigenvector.
Hence, the estimate $\hat {\mathbf h}$ is chosen out of the set $\{(\bm \vartheta_\ell, \lambda_\ell)\}$ of eigenvectors with corresponding eigenvalues of $\hat { \mathbf C }$ and scaled accordingly:
\[
    \ell_\mathrm{princ} = \argmax_\ell \{ \lambda_\ell \} ~~ \rightarrow ~~ \mathbf { \hat h } = \sqrt{\lambda_{\ell_\mathrm{princ}} - \sigma^2} ~ \frac{\bm \vartheta_{\ell_\mathrm{princ}}}{\lVert \bm \vartheta_{\ell_\mathrm{princ}} \rVert}
\]

\subsection{Phase and Power Calibration}
For phase calibration, we again consider the case of a single subcarrier $n$ with $\hat {\mathbf h}_{\mathrm{OTA}} = \hat {\mathbf h}^{(n)}_{\mathrm{OTA}} \in \mathbb C^M$ denoting the estimated channel vector for the over-the-air channel between ESPARGOS antenna array and some WiFi transmitter.

Two types of phase calibrations need to be carried out to achieve phase-coherent receivers:
\begin{itemize}
    \item Removal of the unknown \ac{PLL} / analog hardware phase offsets with the help of the phase reference signal and
    \item calibration for the propagation delays of the phase reference signal passing through the distribution network on the circuit board itself, which manifest themselves in antenna-specific phase offsets $\varphi_{\mathrm{path}, m}$.
\end{itemize}

The \ac{PLL} phase offsets are estimated based on the phase reference signal, resulting in another channel vector estimate $\hat {\mathbf h}_\mathrm{REF} \in \mathbb C^M$.
The estimation procedures for both the over-the-air channel and for the \ac{PLL} phase offsets are carried out independently, as explained in Section \ref{sec:csiacquisition} -- the two different channels (over-the-air and reference signal distribution network) can be distinguished either by the state of the RF switch, or, more reliably, by embedding some reference indicator into the WiFi frames of the reference signal.
By assuming the difference in attenuation across the individual paths of the reference signal distribution network is negligible, the reference signal amplitudes may also be used for power calibration.
Overall, power- and phase-calibrated channel coefficients $\hat h_{\mathrm{CAL}, m}$ which account for both \ac{PLL} / analog hardware phase offsets and reference distribution network phase offsets, can be computed as
\[
    \hat h_{\mathrm{CAL}, m} = \frac{\hat h_{\mathrm{OTA}, m}}{\hat h_{\mathrm{REF}, m}} ~ \mathrm e^{-j \varphi_{\mathrm{path}, m}}.
\]

For a single $2 \times 4$ ESPARGOS antenna array, the antenna-specific phase offsets due to the distribution network path delays $\varphi_{\mathrm{path}, m}$ are a result of the circuit board layout and are constant.
If multiple ESPARGOS antenna arrays are combined, the phase offsets induced by the reference signal distribution network outside of the antenna arrays, typically consisting of power splitters and coaxial cables, also need to be taken into account.

\section{Analyses and Measurements}
\label{sec:measurements}

\subsection{Phase Stability}
\begin{figure}
    \centering
    \includegraphics[width=0.8\columnwidth]{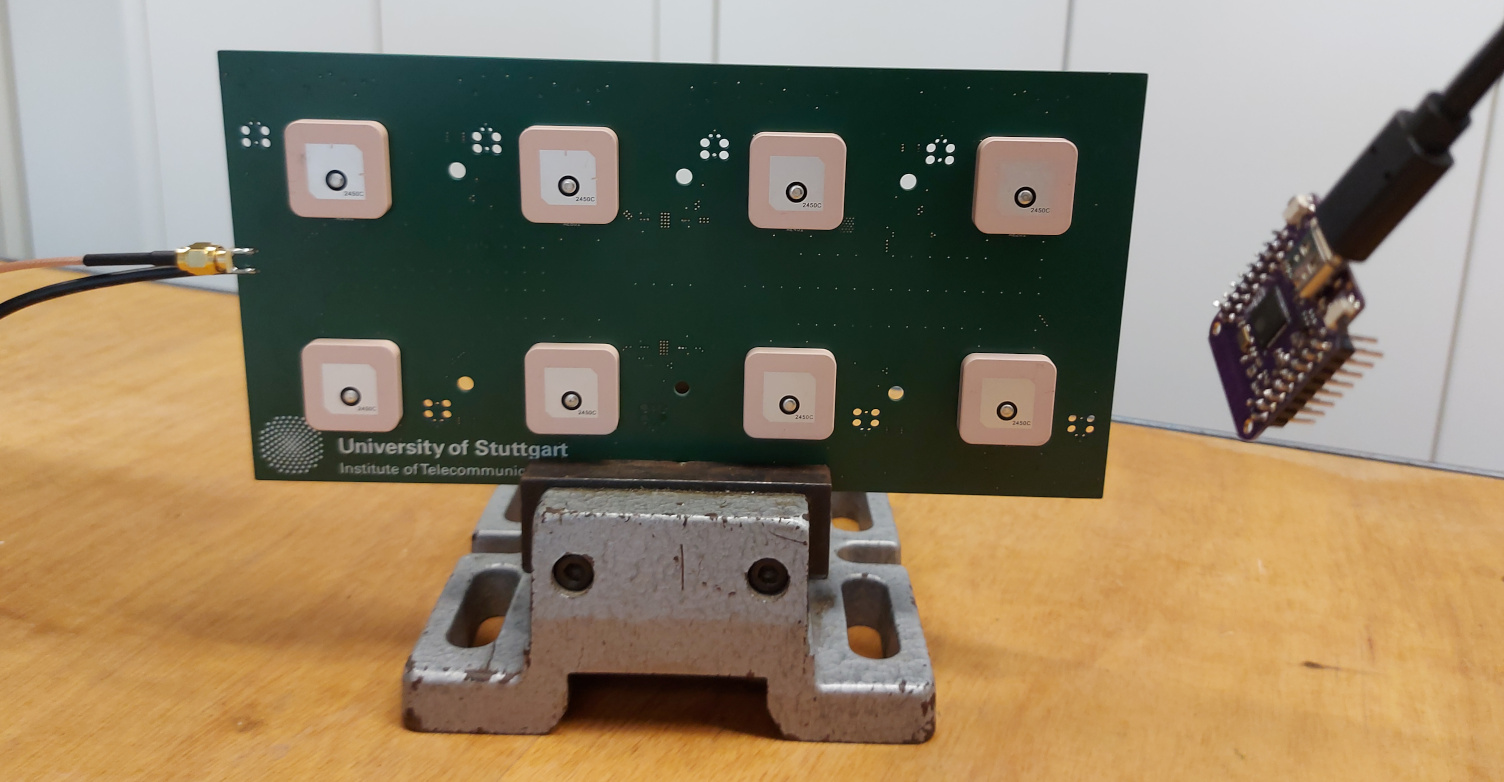}
    \caption{Experiment setup: Frontal view of ESPARGOS and another ESP32 development board used as transmitter}
    \label{fig:frontalview}
\end{figure}

\begin{figure}
    \centering
    \includegraphics[width=0.9\columnwidth]{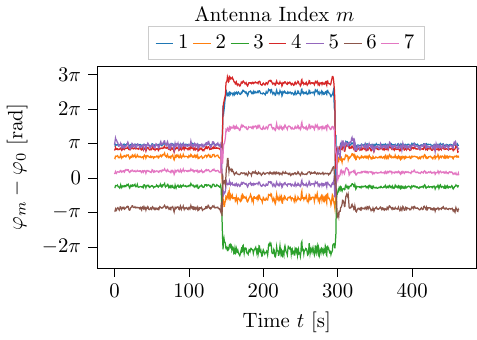}
    \caption{Phase differences (moving average over 20 samples) between antennas over time: Transmitter is relocated at $t \approx 150\,\mathrm{s}$ and moved back to its original location at $t \approx 300\,\mathrm{s}$.}
    \label{fig:phasestability}
\end{figure}

For the phase stability experiment, we forgo aforementioned phase and power calibration step and only take into account the over-the-air channel estimates $\hat h^{(n)}_{\mathrm{OTA}, m}$.
We compute the phase over the sum of the channel coefficients of all $N$ subcarriers:
\[
    \varphi_m = \arg \left\{ \sum_{n = 1}^N \hat h^{(n)}_{\mathrm{OTA}, m} \right\}
\]
Note that this averaging is only sensible for sufficiently frequency-flat channels such as indoor channels with low delay spread.
Long-term phase stability is achieved if, in a static environment, the phase differences $\varphi_m - \varphi_{m_0}$ between any two antennas $m, m_0, ~ m \neq m_0$ remain constant over long intervals.
We typically express all phase differences relative to the same antenna with index $m_0$ which we call the ``reference antenna''.

Synchronizing receivers in frequency using the $40\,\mathrm{MHz}$ reference clock signal should be sufficient for achieving long-term phase stability, even if no phase reference signal is provided.
To test this assumption, a simple experiment was conceived and carried out.
The experiment setup is shown in Fig. \ref{fig:frontalview}:
The ESPARGOS circuit board is fixated to a stand and another, mobile ESP32 microcontroller is used as a WiFi transmitter.
The transmitter is first placed at some marked location, after some time relocated to another place and later moved back to its original location.
Fig. \ref{fig:phasestability}, which shows the resulting measured phase differences over time with reference antenna $m_0 = 0$ for a single $20\,\mathrm{MHz}$ wide WiFi channel, qualitatively demonstrates that phase stability is the case for ESPARGOS:
Unless the transmitter is moved, the variance in the phase difference between any two of the eight antennas in the array is small.

\subsection{Angle of Arrival Measurement}
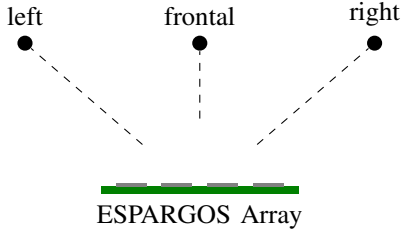
\begin{figure}
    \centering
    \begin{tikzpicture}
        \fill [fill=green!50!black] (0, 0) rectangle (2.6, 0.1);
        \fill [fill=white!50!black] (0.2, 0.1) rectangle (0.6, 0.15);
        \fill [fill=white!50!black] (0.8, 0.1) rectangle (1.2, 0.15);
        \fill [fill=white!50!black] (1.4, 0.1) rectangle (1.8, 0.15);
        \fill [fill=white!50!black] (2.0, 0.1) rectangle (2.4, 0.15);
        
        \node (left) at (-1, 2) [circle, fill = black, inner sep = 2pt] {};
        \node [above = 0.0cm of left] {left};
        \node (frontal) at (1.3, 2) [circle, fill = black, inner sep = 2pt] {};
        \node [above = 0.0cm of frontal] {frontal};
        \node (right) at (3.6, 2) [circle, fill = black, inner sep = 2pt] {};
        \node [above = 0.0cm of right] {right};
        
        \draw [dashed, shorten <=1cm] (1.3, 0) -- (left);
        \draw [dashed, shorten <=1cm] (1.3, 0) -- (frontal);
        \draw [dashed, shorten <=1cm] (1.3, 0) -- (right);
        
        \node [align = center] at (1.3, -0.3) {ESPARGOS Array};
    \end{tikzpicture}
    \caption{Illustration of three considered transmitter placements ``left'', ``frontal'' and ``right'' (not drawn to scale)}
    \label{fig:txplacement}
\end{figure}

In another experiment, the suitability of ESPARGOS for a simple \ac{AoA} estimation task is examined.
Phase and power calibration are applied as previously described and the \ac{AoA} estimates are computed based on the fully calibrated channel coefficient estimates $\hat h_{\mathrm{CAL}, m}$.

\begin{figure}
    \centering
    \begin{subfigure}{0.4\textwidth}
        \centering
        \includegraphics[width=\textwidth]{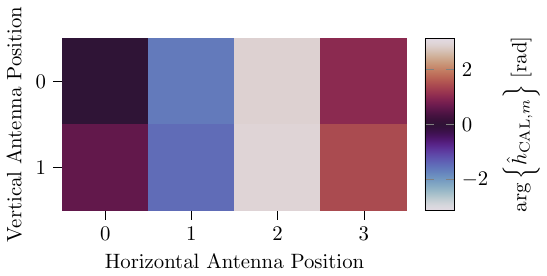}
        \caption{TX to the left of the array}
    \end{subfigure}
    \begin{subfigure}{0.4\textwidth}
        \centering
        \includegraphics[width=\textwidth]{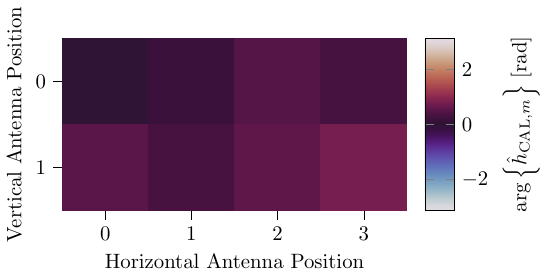}
        \caption{TX in front of the array}
    \end{subfigure}
    \begin{subfigure}{0.4\textwidth}
        \centering
        \includegraphics[width=\textwidth]{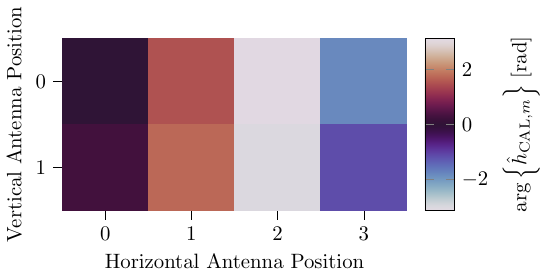}
        \caption{TX to the right of the array}
    \end{subfigure}
    \caption{Received signal phase visualized over antennas, frontal view of ESPARGOS, with transmitter located at different positions relative to the receiver array. For the illustration, the top-left antenna in the array $m_0 = 0$ (vertical and horizontal index 0) is used as a phase reference (i.e., its phase is always $\arg \left\{ \hat h_{\mathrm{CAL}, 0} \right\} = 0$).}
    \label{fig:receivedphases}
\end{figure}

As a proof for the feasibility of the system, we only consider a simple qualitative test, namely whether ESPARGOS can distinguish between three different transmitter placements.
In this setup, ESPARGOS is again fixated in a stand as shown in Fig. \ref{fig:frontalview} and channel coefficients for the three transmitter locations illustrated in Fig. \ref{fig:txplacement} are measured.
Namely, the transmitter is either located directly in front of the antenna array at a distance of a few meters, or shifted to the right or to the left from the frontal position.

The phases of the estimated channel coefficients $\hat h_{\mathrm{CAL}, m}$ for the three different transmitter locations are illustrated in Fig. \ref{fig:receivedphases}, by coloring antennas according to the received phases in frontal views of the antenna array.
Even by just looking at the measured phase patterns, it is easy to see that information about the \ac{AoA} is contained within $\hat h_{\mathrm{CAL}, m}$:
Depending on the transmitter location, the received phases either increase from left to right, from right to left or are approximately the same over all antennas.

\begin{figure}
    \centering
    \begin{subfigure}{0.39\textwidth}
        \centering
        \includegraphics[width=0.9\textwidth]{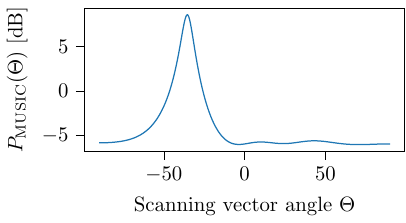}
        \caption{TX to the left of the array}
    \end{subfigure}
    \begin{subfigure}{0.39\textwidth}
        \centering
        \includegraphics[width=0.9\textwidth]{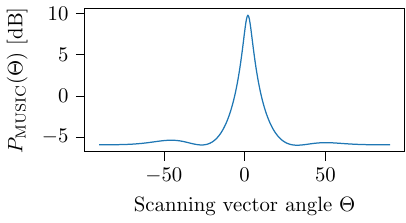}
        \caption{TX in front of the array}
    \end{subfigure}
    \begin{subfigure}{0.39\textwidth}
        \centering
        \includegraphics[width=0.9\textwidth]{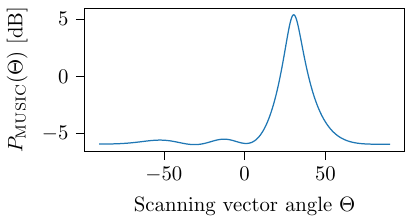}
        \caption{TX to the right of the array}
    \end{subfigure}
    \caption{Spatial pseudo-spectra computed with the \ac{MUSIC} algorithm}
    \label{fig:spatialspectra}
    \vspace{-0.4cm}
\end{figure}

For a practical application of ESPARGOS, \ac{AoA} estimatation algorithms such as \ac{MUSIC} \cite{schmidt1986multiple} could be applied to the received channel coefficients $\hat h_{\mathrm{CAL}, m}$.
As explained in \cite{schmidt1986multiple}, \ac{MUSIC} computes a spatial pseudo-spectrum by assigning a power value to every potential angle $\Theta$ of an incident wave.
Clearly, the spatial pseudo-spectra obtained for the measurements using the \ac{MUSIC} algorithm, shown in Fig. \ref{fig:spatialspectra}, easily permit distinguishing between the three different transmitter locations.

\section{Conclusion and Outlook}
\label{sec:conclusion}
The concept of a realtime-capable WiFi channel sounder which, despite its very low cost, can achieve a high degree of spatial diversity was presented and a prototype was qualitatively evaluated in a simple practical application.
Quantitative performance testing and further characterizations of the system remain a subject for future study.
While ESPARGOS could never replace dedicated multi-antenna channel sounders employed in wireless research, it could be a useful tool for experiments in domains like \ac{JCaS} and security research if the quality of measurements turns out to be sufficient for such applications.
In the future, we plan to apply ESPARGOS to applications like Channel Charting.
Furthermore, the basic concepts behind the architecture of ESPARGOS are transferable to similar application-specific channel sounders.

\bibliographystyle{IEEEtran}
\bibliography{IEEEabrv,references}

\end{document}